\documentclass[12pt,preprint]{aastex}

\newcommand\msun{\rm M_{\odot}}

\newcommand\rsun{\rm R_{\odot}}
\newcommand\msunyr{\rm M_{\odot}\,yr^{-1}}
\newcommand\be{\begin{equation}}
\newcommand\en{\end{equation}}

\newcommand\etal{{\rm et al}.\ }
\newcommand\haebe{\rm HAe/Be}
\newcommand\mdot{\dot{M}}
\newcommand\Md{\dot{M}}
\newcommand\curf{{\cal F}}
\newcommand\ecs{\rm erg \, cm^{-2} \, s^{-1}}

\begin{document}
\title{Magnetospheres and Disk Accretion 
in Herbig Ae/Be Stars}
\author{
James Muzerolle\altaffilmark{1,2},
Paola D'Alessio\altaffilmark{3},
Nuria Calvet\altaffilmark{4},
and Lee Hartmann\altaffilmark{4}}
\altaffiltext{1}{Steward Observatory, 933 N. Cherry Ave., University of Arizona,
Tucson, AZ 85721}
\altaffiltext{2}{Visiting Astronomer, Kitt Peak National Observatory,
National Optical Astronomy Observatory, which is operated
by the Association of Universities for Research in Astronomy, Inc.
(AURA) under cooperative agreement with the National Science Foundation.}
\altaffiltext{3}{Instituto de Astronom\'{\i}a, UNAM, Ap. Postal
3-72, 58089 Morelia, M\'exico}
\altaffiltext{4}{Harvard-Smithsonian Center for Astrophysics, 60
Garden St., Cambridge, MA 02138}

\begin{abstract}
We present evidence of magnetically-mediated disk accretion in Herbig Ae/Be
stars.  Magnetospheric accretion models of Balmer and sodium profiles
calculated with appropriate stellar and rotational parameters are
in qualitative agreement with the observed
profiles of the Herbig Ae star UX Ori, and yield a mass accretion rate of
$\sim 10^{-8} \; \msunyr$.  If more recent indications of an extremely
large rotation rate for this object are correct, the magnetic field
geometry must deviate from that of a standard dipole in order to produce
line emission consistent with observed flux levels.
Models of the associated accretion shock qualitatively
explain the observed distribution of excess fluxes in the Balmer discontinuity
for a large ensemble of Herbig Ae/Be stars, and imply typically small mass accretion
rates, $\lesssim 10^{-7} \; \msunyr$.  In order for accretion to proceed onto
the star, significant amounts of gas must exist inside the dust destruction radius,
which is potentially problematic for recently advocated scenarios of ``puffed"
inner dust wall geometries.  However, our models of the inner gas disk show
that for the typical accretion rates we have derived, the gas should be
generally optically thin, thus allowing direct stellar irradiation of the inner
dust edge of the disk.
\end{abstract}


\section{Introduction}

The class of Herbig Ae/Be ($\haebe$) stars was identified by Herbig (1960)
in an attempt
to find the analogues of T Tauri stars among objects of higher mass.
This identification has been supported by many studies over the last several years.
Observations of mm-wave emission suggest the presence of circumstellar
disks similar to those of the low-mass classical T Tauri stars (CTTSs)
(Mannings, Koerner, \& Sargent 1997; Mannings \& Sargent 1997; Testi \etal 2000).
Dusty disk models can in principle explain the infrared excess emission 
of many $\haebe$ stars (Hillenbrand et al. 1992; Chiang \etal 2000;
Natta \etal 2001; Dullemond, Dominik, \& Natta 2001, DDN01).  In addition, 
imaging observations in scattered light (e.g., Grady \etal 1999) 
also provide evidence for flattened or disk-like large scale dusty 
structures around a few $\haebe$ stars.

These studies naturally raise the question:
are $\haebe$ stars accreting from their disks?  And if so, are the accretion
rates high enough to modify stellar evolution, or to imply substantial
migration of potential planet-forming material? 
The first attempt to derive $\haebe$ accretion rates was made by Hillenbrand
\etal (1992; HSVK92), who attempted to fit the observed infrared excess emission using
steady accretion disk models and estimated much higher mass accretion rates
($10^{-6} - 10^{-5} \; \msunyr$) than typical of CTTSs.  However, HSVK92 also needed
to invoke an inner disk hole, of order $0.1$~AU in radius, to explain the
decline in near-infrared excess emission shortward of $\lambda \sim 2-3 \mu$m.
Hartmann, Kenyon, \& Calvet (1993; HKC) argued that the accretion could not
stop indefinitely at this inner radius, and showed that 
accretion at the rates inferred by HSVK92 would render any inner disk 
optically thick, eliminating the decline in infrared excess at short wavelengths.

Natta \etal (2001) and DDN01 revisited this problem,
this time developing models which explain the infrared excess emission
of $\haebe$ stars as the result of heating by irradiation from the central star,
rather than local accretion energy release.  These authors again assumed
an inner disk hole to explain the decrease in near-infrared disk emission
at short wavelengths, of size $\sim 0.3 - 0.5$~AU.  The inner edge of the disk
receives radiation from the star at near-normal incidence, and therefore becomes
much hotter at a given radius than would be a geometrically thin, 
optically thick disk, irradiated obliquely.  The increased heating 
thus causes the disk material at the inner edge to
``puff up''.  DDN01 suggested that this expanded inner disk edge 
can quantitatively explain the magnitude of the near-infrared emission in $\haebe$ 
spectral energy distributions (SEDs), without any accretion energy release.
The location of the inner disk edge and its emission in this model
is roughly consistent with recent interferometric observations
(Millan-Gabet, Schloerb, \& Traub 2001), which indicate near-infrared
H and K band emission at radial distances well beyond that predicted by
standard geometrically thin, optically thick, irradiated disks.

While this model is attractive, the possibility of accreting material interior to
the dust destruction radius must be addressed.
All the low-mass T Tauri stars with near-infrared excesses exhibit accretion
onto the central star, and it would be surprising if HAe/Be disks with similar
properties were completely inert.  Indeed, heating of the inner disk edge 
by stellar radiation alone is likely to raise temperatures to the point
that the fractional ionization is high enough that the magnetorotational instability
(MRI)
thought to drive disk accretion can operate (e.g., Balbus \& Hawley 1998; Gammie 1996).
In CTTSs, stellar magnetospheres truncate the inner disk, but as pointed out by HKC, 
the inner disk edge in the DDN01 model lies at such large radii that
magnetospheric truncation is implausible, and in any event falls
well outside of the corotation radius (where the disk Keplerian angular
velocity matches the stellar angular velocity), which would therefore prevent
accretion (e.g., Shu \etal 1994).  Beyond this, there is direct
evidence for ballistic infall in some $\haebe$ stars (\S 2; Sorelli, Grinin,
\& Natta 1996; Natta, Grinin, \& Tambovtseva 2000), which implies that material
must extend well inward of the dust destruction radius. 
DDN01 argue that their model will hold even if accretion proceeds
through the dust destruction radius, as long as the accreting gas is optically
thin, but this assumption places severe constraints on accretion parameters
(HKC).  Are accretion rates low enough that the inner accretion disk is optically thin?  
Or is there some other explanation of the near-infrared SED?

In this paper we investigate accretion in $\haebe$ systems
and its implications for inner disk structure.
We begin by analyzing a specific object of special interest, UX Ori,
in terms of a magnetospheric accretion model similar to that developed
by Muzerolle \etal (1998, 2001).  Next, we consider possible diagnostics
of accretion from optical-ultraviolet continuum emission; these diagnostics
limit accretion to rates much lower than those typically assumed by HSVK92.  
Finally, we investigate the inner disk structure in these systems.  We show
that for low accretion rates, the inner disk is likely to be optically thin,
as assumed by the Natta \etal (2001) and DDN01 models; for somewhat higher
accretion rates, the innermost disk may be optically thick but geometrically
thin, allowing the direct irradiation of most of the ``dust edge'' of the disk.
In this way we attempt to construct a more complete picture of $\haebe$ systems. 

\section{Accretion rates}

An important constraint on the optical depth in an accretion disk
is the mass accretion rate $\mdot$;  higher $\mdot$ generally leads
to greater optical depths in the disk (HKC).  In this section we
consider constraints on $\mdot$ from line profiles and UV/optical veiling,
in the context of magnetospheric accretion.  While early-type stars are thought
to  have much weaker magnetic fields due to dynamo activity than their
cooler pre-main sequence counterparts, some main sequence
A stars do exhibit ordered magnetic fields of strengths $\sim$ kG 
(Mestel 1975; Borra, Landstreet, \& Mestel 1982), which would
be sufficient to truncate inner disks at a few stellar radii
(though not at the inner disk radii $\sim 0.1$~AU required by
the DDN01 models; Hartmann 1999).  

Moreover, magnetospheric accretion
models are ideal for explaining high-velocity redshifted absorption 
in some $\haebe$ stars (Sorelli \etal 1996, and references therein).  
An alternate explanation of the ubiquitous 
redshifted absorption features seen in a number of $\haebe$ stars 
is infall of cometary bodies (e.g., Grinin \etal 1994, G94;
Sorelli et al. 1996; Grady et al. 1999).
In this scenario, star-grazing planetesimals will evaporate,
and, given a favorable viewing angle, redshifted absorption
of neutral sodium and other atoms with low ionization potentials
can result.  This hypothesis provides a potential link to
the older $\beta$ Pic objects with debris disks.  However,
a separate source, such as a wind or accretion flow is required
to explain the line emission observed in stars such
as UX Ori.  Moreover, the infall signatures,
while variable, are not episodic as in $\beta$ Pic
(e.g., Lagrange, Backman, \& Artymowicz 2000, and references therein).
Most significantly, Natta et al. (2000) found that the infalling
gas is not strongly depleted in hydrogen, contrary to expectation
if the source were evaporating comets (note that no
circumstellar hydrogen absorption or emission has been
observed in $\beta$ Pic; Lagrange, Backman, \& Artymowicz 2000).

Below we consider magnetospheric accretion in more detail by modeling
line profiles of one well-studied object, UX Ori.  This star 
is particularly interesting for its large-amplitude optical variability, 
probably due to time-variable extinction by dust clouds
(Grinin \etal 1994; Natta \& Whitney 2000).

\subsection{Line profiles}

\begin{figure}
\plotone{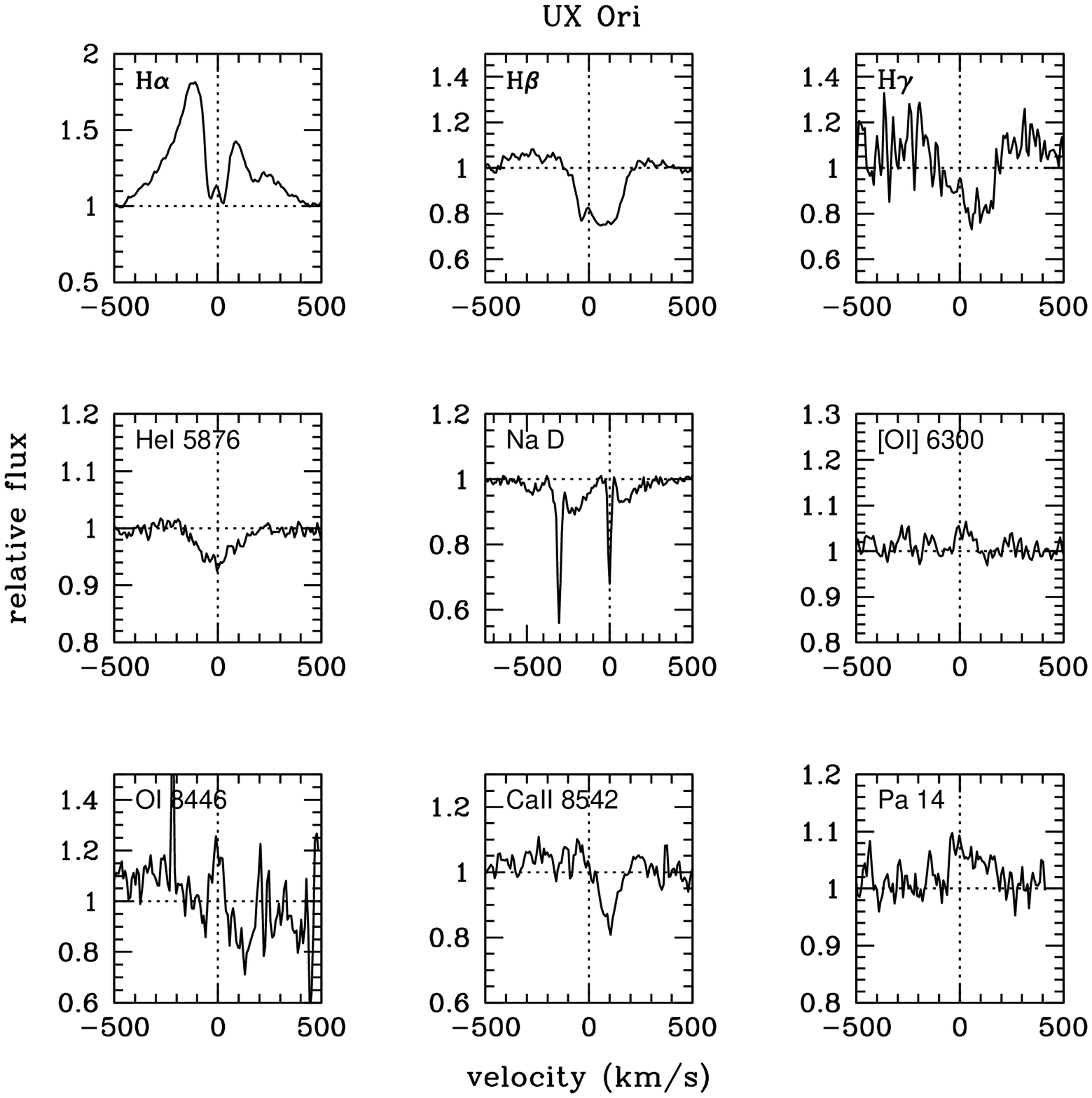}
\caption{Line profiles from our echelle spectrum of UX Ori.
Each profile was normalized by the continuum, and then subtracted
by an A2 standard star spectrum.
\label{obsprof}}
\end{figure}

UX Ori was observed with the echelle spectrograph on the 4m
telescope at KPNO on September 17, 1999. 
The spectrum has a resolution of $\sim30,000$,
and a wavelength coverage of
4300 - 9200 {\AA}, containing many features of interest such as
the Balmer lines, He I, Na D, and the Ca II infrared triplet.
The data were reduced using the standard IRAF echelle packages.
In order to provide pure emission profiles, the spectrum of
standard HD 1280 (spectral type A2, $v\sin i=90-100$ km s$^{-1}$)
was subtracted from the
object spectrum, effectively removing photospheric absorption
features (including the Balmer absorption wings).
Figure~\ref{obsprof} shows the continuum-normalized profiles
of nine atomic lines.  
The observed line profiles for UX Ori exhibit characteristics
similar to those seen before 
(G94; Natta et al. 2000; Grinin \etal 2001).  Few lines are in emission,
but many exhibit redshifted absorption at supersonic velocities
as high as several hundred km s$^{-1}$
(e.g., Balmer, Na D, Ca II), indicating mass infall onto the star.  

We constructed magnetospheric accretion models for comparison
with the observed profiles of UX Ori.
The radiative transfer models are described in detail in
Muzerolle et al. (2001).  In summary, we assume a dipolar
magnetic field geometry; gas accreting from a circumstellar
disk falls ballistically along the field lines to the stellar
surface.  The density of the infalling gas is set by the
geometry, velocity, and parameterized accretion rate.
The inner and outer radius of the flow are free parameters
(though the outer radius must be less than the disk corotation
radius).  Finally, we use a solid-body treatment for the
rotation of the magnetosphere, including curvature
of the field lines in the direction of rotation (Muzerolle et al. 2001).

Model line profiles were calculated in the same
manner as for the T Tauri comparisons with rotation,
using the same gas temperature constraints found in Muzerolle
et al. (2001) (i.e., in the range 6000-12,000 K, varying inversely
with the density).  Corresponding to the observed properties
of UX Ori, an A2 star, the model stellar parameters were changed
to $M_* = 3 \, \msun$ and $R_* = 3 \, \rsun$.
We investigated two different values for the stellar rotation
velocity, 70 and 140 km s$^{-1}$, following $v\sin i$ measurements
from B\"ohm \& Catala (1995) and Grinin et al. (2001), respectively,
and given the expected edge-on orientation of the system.
A more recent determination found $v\sin i = 215$ km s$^{-1}$
(Mora et al. 2002); the reason for these hugely discrepant values
is unclear.  We discuss the effects of rotation and its implications
on the size of the accretion flow below.

\begin{figure}
\plotone{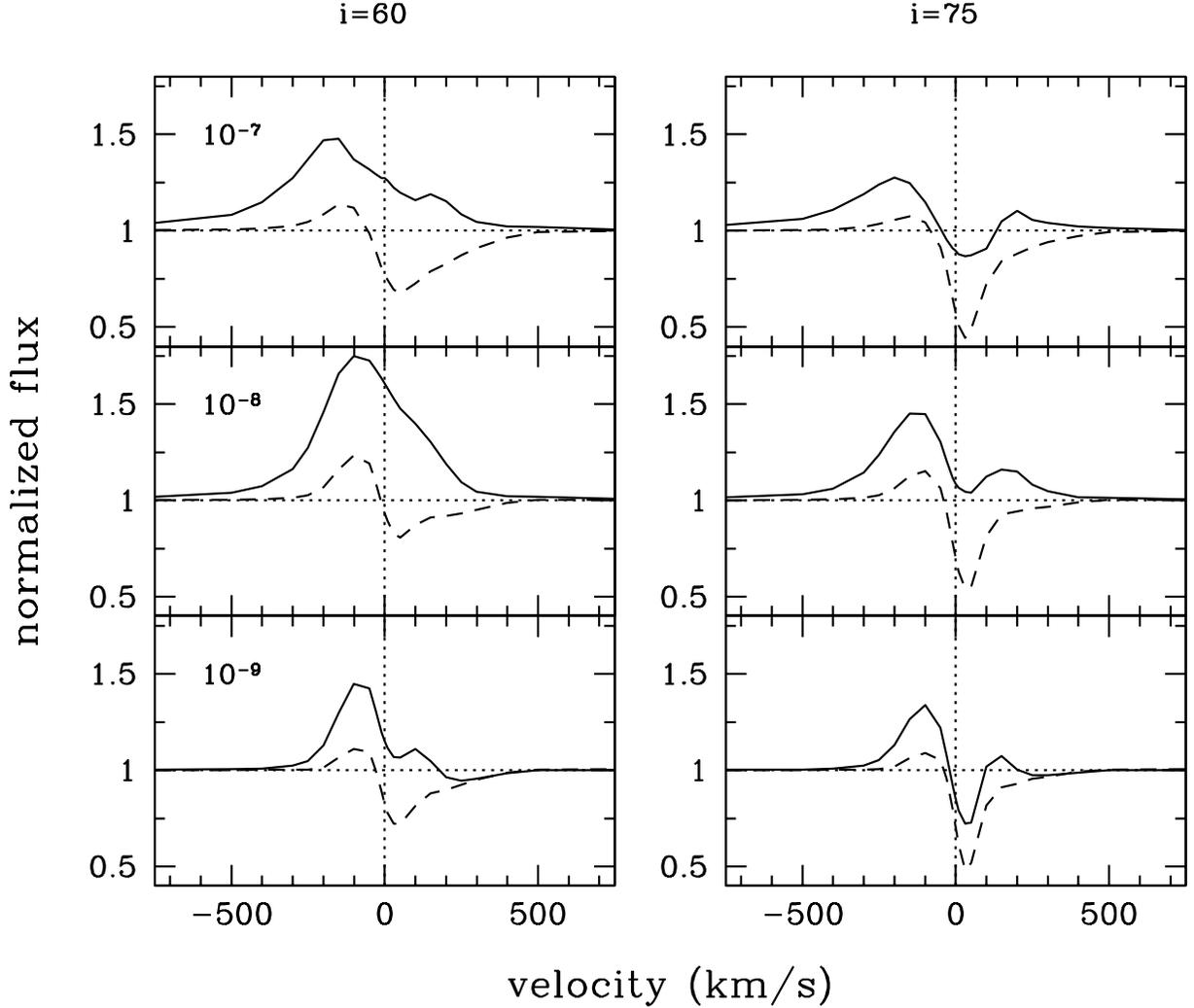}
\caption{Magnetospheric accretion model line profiles of H$\alpha$ (solid
line) and H$\beta$ (dashed line) for the indicated inclination angles
(in degrees, with zero
being pole-on and 90 edge-on) and values of $\mdot$ (in units of
$\msunyr$).  All models were calculated with $M=3 \, \msun$,
$R=3 \, \rsun$, $V_*=70$ km s$^{-1}$, and magnetospheric radii
$R_{mag}=2.4-2.9$.
Maximum gas temperatures were $T_{max}=8000$ K for $\mdot = 10^{-7}
\; \msunyr$, 10,000 K for $\mdot = 10^{-8} \; \msunyr$, and 12,000 K
for $\mdot = 10^{-9} \; \msunyr$.
\label{mod_mdot}}
\end{figure}

Figure~\ref{mod_mdot} shows a small sample of UX Ori model profiles,
demonstrating the general
effects of the accretion rate and inclination angle on the Balmer
line profile shape and flux.  Higher accretion rates produce
broader emission wings from opacity broadening, as well as
broader absorption from the increasing continuum opacity,
which puts a limit on the peak flux such that the largest
line fluxes occur at an intermediate accretion
rate of $\sim 10^{-8} \; \msunyr$.  Larger inclination angles
result in more significant line asymmetries, including the
appearance of broad, low-velocity redshifted absorption
produced by the occultation of the hot star by the cooler
outer parts of the magnetosphere.  This particular feature,
along with the stronger blue emission peak, is almost always
seen in H$\alpha$ profiles of UX Ori and similar objects,
{\it and can only be explained by infalling material}.

The constraints our models can ultimately place on the accretion rate
are limited by uncertain constraints on other model parameters,
especially the gas temperature.  For example, in some cases
the same emission levels can be reproduced by increasing the
temperature and decreasing the density (i.e., $\mdot$),
or vice-versa, beyond the T Tauri temperature constraints.
However, we find that $\mdot \gtrsim 10^{-7} \; \msunyr$ cannot
reproduce the observations at any temperature: at $T \gtrsim 8000$ K,
the gas continuum optical depth becomes significant,
resulting in broad line absorption at H$\alpha$,
while at lower temperatures,
the H$\alpha$ emission is much weaker than observed.
On the other hand, values of $\mdot$ much lower than
$10^{-8} \; \msunyr$ also cannot reproduce the observed H$\alpha$
emission, especially in the wings.  In that case, simply increasing
the gas temperature above that of the adopted constraints
cannot result in much more line flux since the gas is
already almost completely ionized.  Ultimately, we believe
a model-derived mass accretion rate estimate should be accurate
to within about a factor of 5.

\begin{figure}
\plotone{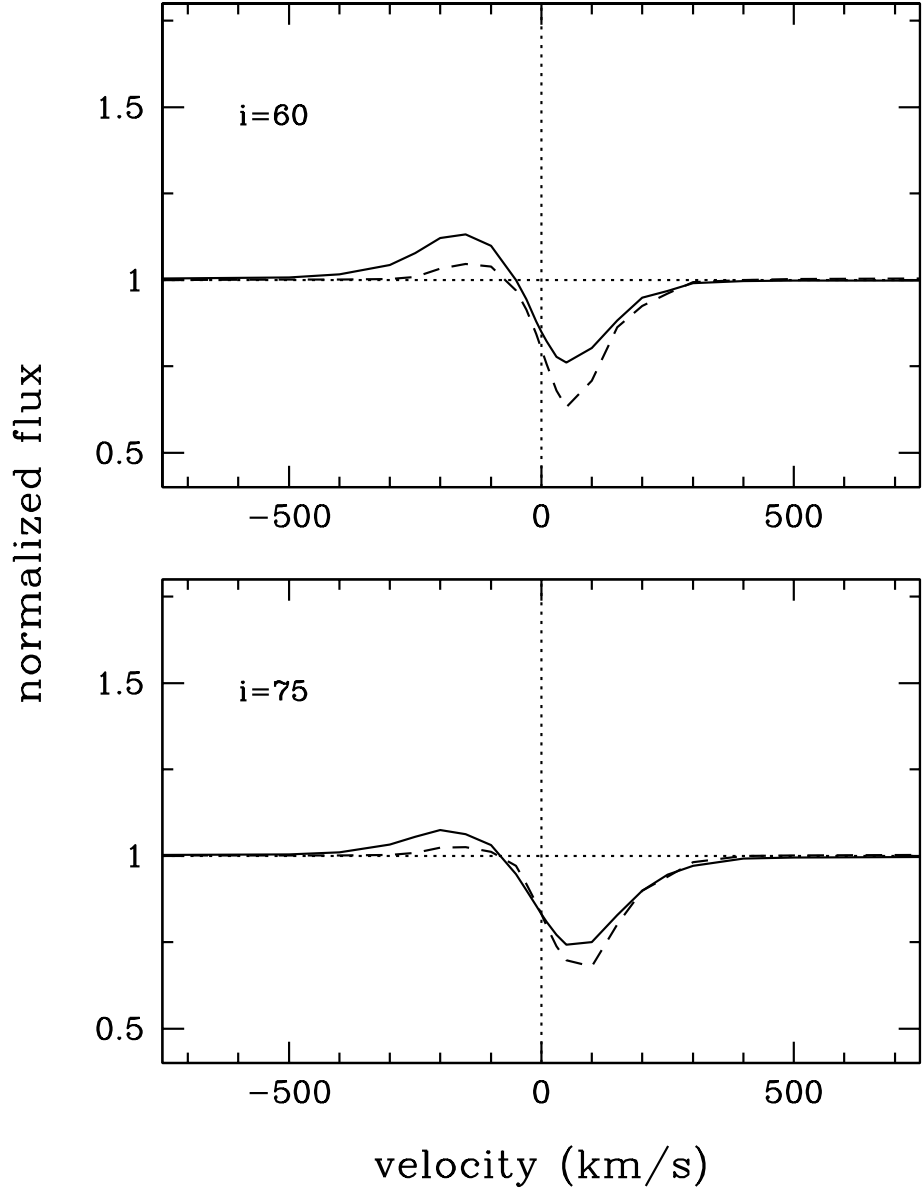}
\caption{H$\alpha$ (solid line) and H$\beta$ (dashed line) model
line profiles for a smaller magnetosphere, with
$R_{mag}=1.35-1.85$, corresponding to a stellar rotation rate
of $V_* = 140$ km s$^{-1}$.  $\mdot = 10^{-8} \; \msunyr$,
$T_{max}= 10,000 K$.
\label{mod_small}}
\end{figure}

\begin{figure}
\plotone{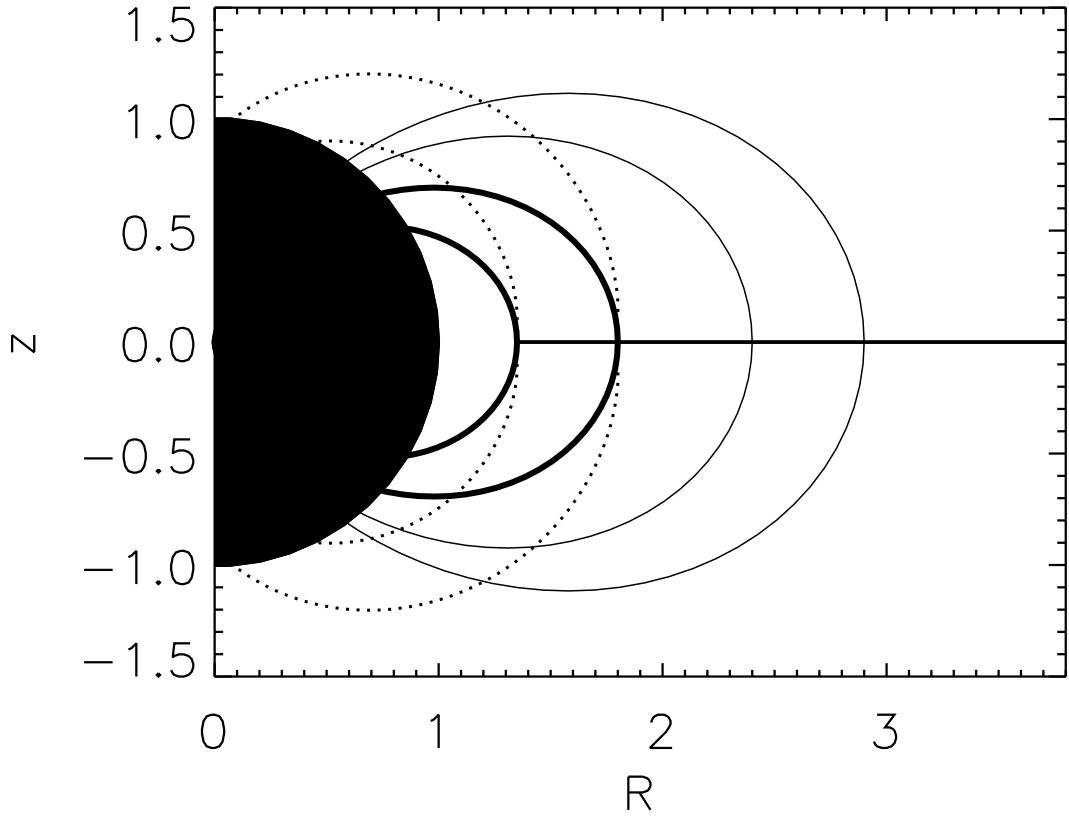}
\caption{Comparison of different magnetosphere sizes (light and dark
solid lines) used for the previous two figures,
and an approximation of a ``pinched" dipole configuration (dotted line),
as discussed in the text.
\label{geom}}
\end{figure}

As already mentioned, another difficulty is the stellar rotation
velocity, and the resulting limit on the size of the accretion flow.
We find that the higher values of $v\sin i$ are problematic for the
magnetospheric accretion model in that they result in
very small corotation radii (2.1 and 1.6 $R_*$, respectively,
for $v\sin i = 140, 215$ km s$^{-1}$),
and, hence, extremely small accretion flows since accretion
cannot take place outside of corotation.  For such small
magnetospheric radii, there is not enough emission volume
to match the observed H$\alpha$ emission with any other combination
of parameters; an example is shown in Figure~\ref{mod_small}.
We obtain the best match to the observations using a larger magnetospheric
radius, 3 $R_*$, roughly equivalent to the corotation radius for
$V_* \sim 70$ km s$^{-1}$.  We note that
the actual geometry of any magnetic fields that may be present in
Herbig Ae stars is completely unknown.  A ``pinched" or nonaxisymmetric
geometry that deviates from a standard dipole may allow for a model
with a higher rotation velocity that has a larger emission volume
and can reproduce the observations (Fig.~\ref{geom} shows one
possibility).  Calculating the radiative transfer for such a
complicated geometry, however, is beyond the scope of this paper.

\begin{figure}
\plotone{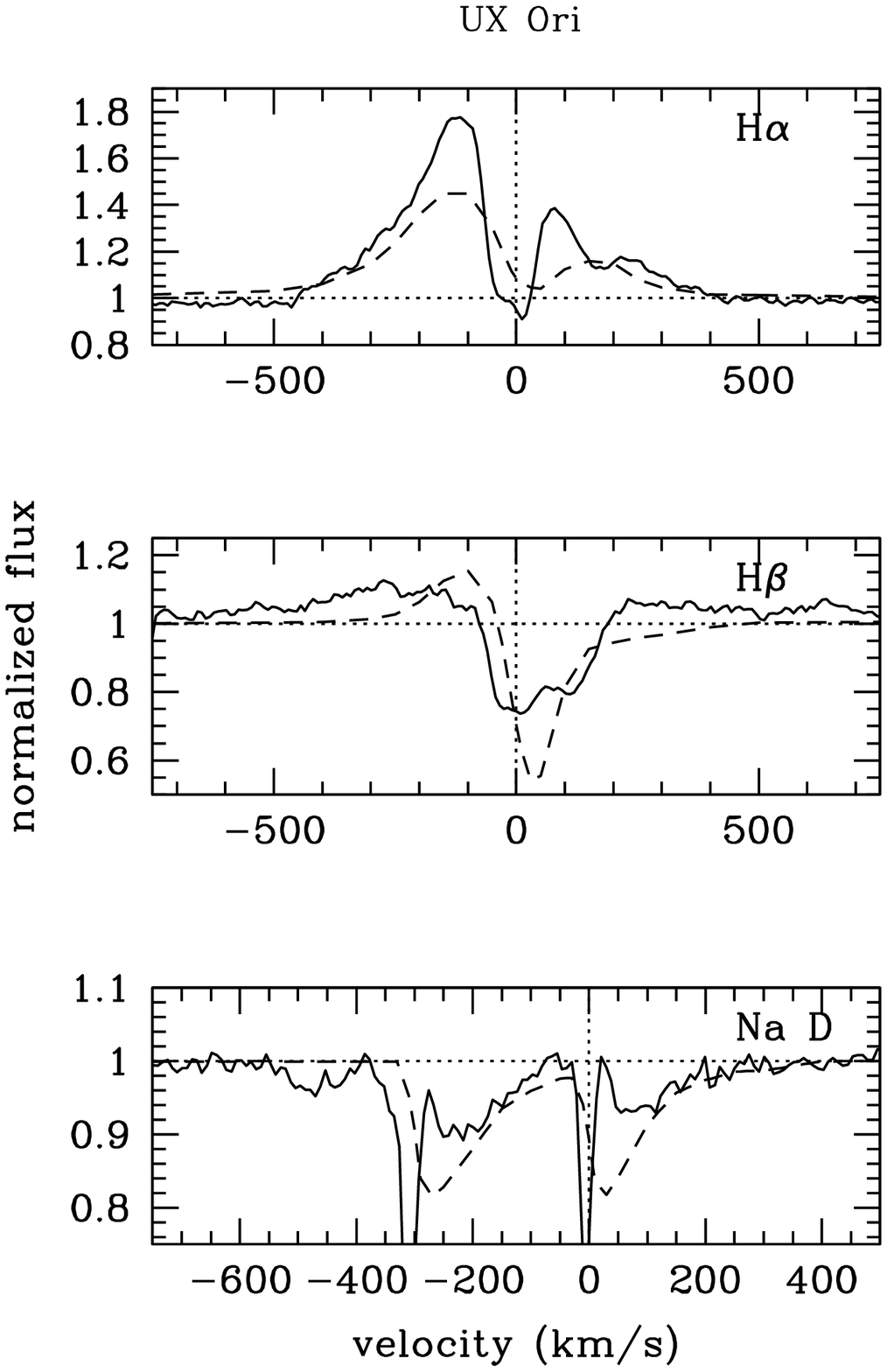}
\caption{Observed (solid lines) and model (dashed lines)
line profiles for UX Ori.  The model parameters are:
$\mdot=10^{-8} \; \msunyr$; $i=75^\circ$;
$R_{mag}=2.4-2.9 \; \rsun$; $T_{max}=10,000$ K;
$T_{phot}=8000$ K; $V_{rot}=70 \; {\rm km \, s^{-1}}$.
\label{profcomp}}
\end{figure}

Figure~\ref{profcomp} shows the observed and best model match profiles
of H$\alpha$, H$\beta$, and Na D for UX Ori.
The magnetospheric infall models qualitatively match the
observed line profiles in all three cases.
The model accretion rate, $10^{-8} \; \msunyr$, is consistent
with the upper limit we derive from the lack of hot continuum
excess (see next section).  The inclination angle for these models is
$75^\circ$, in agreement with a nearly edge-on
orientation of the system inferred from strong extinction
events (e.g., G94).  There is a central,
slightly blueshifted absorption component in the Balmer lines
that is not fit.  This feature could be due to absorption from an
accretion-driven wind exterior to the emission region (as seen in
most CTTSs), which should be oriented roughly perpendicular to
the accretion disk, resulting in small blueshifted absorption
velocities given the edge-on orientation of the disk.
The central absorption in the sodium lines is the interstellar
component; there may also be some residual contamination
from imperfectly subtracted night sky lines.  The discrepancy
in the level and
velocity of the redshifted absorption components probably
reflects deviations from the idealized flow geometry,
as well as uncertainties in the gas temperature distribution.
In any case, we feel the general agreement is reasonable,
especially given the variable nature of the lines -- note
the similarity of these models to the multi-epoch spectra
of UX Ori in G94.  Similar H$\alpha$ and Na D profiles have also
been observed in spectra of BF Ori
(de Winter et al. 1999), a Herbig Ae star which exhibits
photometric variability similar to that of UX Ori.
Finally, we also note that our accretion rate estimate
is comparable to that derived by
Tambovtseva \etal (2001) for a model in which the infalling
material lies essentially in the equatorial plane.

As noted above, UX Ori is seen to
to dim by up to several magnitudes in the optical (Herbst
et al. 1983, 1994; Bibo \& Th\'e 1990), accompanied by
increased polarization (G94), which is interpreted
as the result of an extinction event.  During one such
minimum, G94 obtained an optical spectrum.
The H$\alpha$ profile exhibited a larger emission equivalent
width (though with a lower total flux), with no absorption
components, and a single emission peak blueshifted by nearly
50 $\rm{km \, s^{-1}}$.  Meanwhile, the redshifted absorption
in the Na D lines almost completely disappeared.  These authors
hypothesized that an optically thick dust cloud obscured
most of the star during the minimum, while at the same time
covering up much of the line emitting region, causing
the change in profile shapes.  Grinin \& Tambovtseva (1995)
calculated H$\alpha$ profiles in rough agreement with
the observations using a model of the circumstellar
obscuration, and assuming gas in the inner disk with a velocity
distribution similar to our magnetospheric accretion models.

\begin{figure}
\plotone{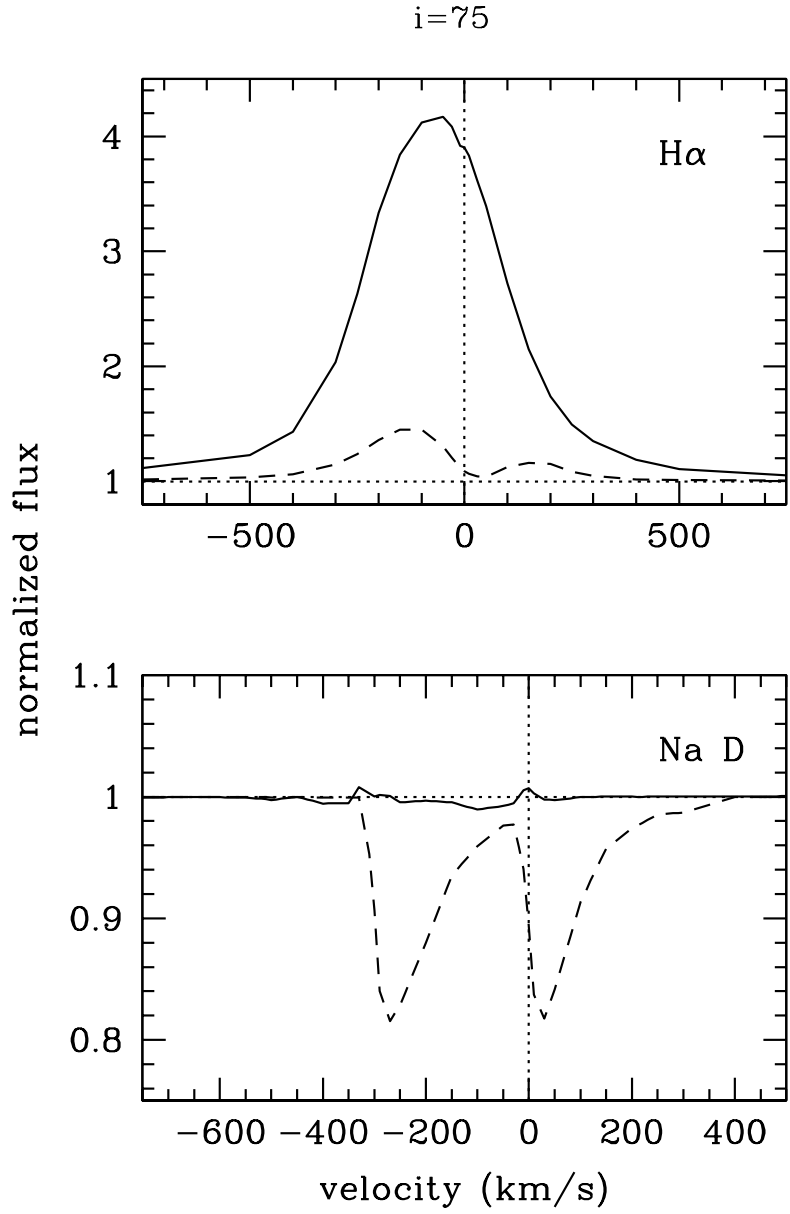}
\caption{Model profiles computed as in Figure~\ref{profcomp}
(dashed lines), compared to calculations with an opaque screen
covering most of the star and emitting region (solid lines).
\label{wallprof}}
\end{figure}

We attempted to simulate this obscuration with our
models, to compare with both H$\alpha$ and Na D.
We included an opaque, nonemitting "screen" extending above and
below the disk to a height of about 2.7 $R_*$,
leaving just a small portion of the star and the magnetosphere
unocculted.  Such an occultation
could be the result of density inhomogeneities or a warp
in the disk or inner dust wall, enhanced by the nearly edge-on orientation
of the system.
The resulting profiles, shown in Figure~\ref{wallprof},
are in good agreement with the observations at photometric
minimum.  H$\alpha$ has a much larger equivalent width,
and shows a single peak blueshifted by about 50-100
$\rm{km s^{-1}}$.  The redshifted absorption in the sodium
lines has completely disappeared.  This behavior is easily
explained as a result of the geometry of the accretion flow.
The part of the flow moving away from the observer covers
much of the star at the inclination angle used here
($75^\circ$); the projection of this gas against the
hot star results in redshifted absorption.  The simulated
obscuration occults most of this region, and the only
part of the flow still observed is that moving towards
the observer from behind the star.  Thus, H$\alpha$ shows
a strong, blueshifted emission peak, and sodium does not
show redshifted absorption (the emission measure of sodium
is not high enough to produce any emission in this case).
The similarity of the profiles in Figure~\ref{wallprof}
to those shown in Figures 4 and 5 of G94
lend further support for magnetospheric accretion taking
place in UX Ori.

\subsection{Magnetospheric accretion shock}

In the magnetospheric accretion model, material
striking the stellar surface merges into the star
through an accretion shock, from which most of the
accretion luminosity is released. Models
for the continuum emission  from the accretion shock 
in CTTSs
have been calculated by Calvet \& Gullbring (1998, CG98).
In these models, the gas in a plane-parallel
accretion column, carrying energy
flux $\curf$, impacts the stellar
surface and shocks to a temperature
$T_s \sim 8.6 \times 10^5 \; (M_*/0.5 \, \msun)/(R_*/2 \, \rsun)$ K,
where $M_*$ and $R_*$ are the stellar mass and radius.
The shock releases soft X-ray radiation which is absorbed by
the accretion stream above and the stellar photosphere below the shock, 
producing optical and UV emission as it thermalizes.
Approximately 3/4 of the emerging total column luminosity 
is emitted by the heated atmosphere
below the shock, while the rest arises in the pre-shock region
(CG98). 

CG98 and Gullbring et al. (2000) show that
the excess continuum flux that veils the photospheric lines
of T Tauri stars in the visible,
and produces fluxes orders of magnitude above
photospheric fluxes in the UV, can be explained
in terms of the emission of accretion columns carrying
energy fluxes of the
order of $\curf \sim 10^{11} - 10^{12} \; \ecs$,
with a small surface coverage of accretion columns,
$f \le$ 1\%, where $f$ is the filling factor.
The high mass accretion rate stars, which
show almost featureless spectra, could
be explained with higher surface coverage,
$f \le $ 10\%.

If $\haebe$ stars undergo magnetospheric accretion, then
an accretion shock must be formed at the
stellar surface.  Here, we calculate accretion
shock models for stellar parameters appropriate to
early A accreting stars, and with energy fluxes
similar to those characterizing the lower
mass accreting stars, following similar
procedures to those in CG98.
The total
flux emerging from the star is
\be
F_{\lambda} = f F_{\lambda}^{col} + ( 1 - f ) F_{\lambda}^{phot}
\label{totalf}
\en
where $F_{\lambda}^{phot}$ is the flux from the
undisturbed photosphere and $ F_{\lambda}^{col} =
F_{\lambda}^{hp} + F _{\lambda}^{pres}$ is
the flux from the accretion columns, which in turn
is the sum of the fluxes from the
heated photosphere, $F_{\lambda}^{hp}$, and the preshock region,
$F _{\lambda}^{pres}$. The parameter $f$ is the surface filling
factor of accretion columns, 
estimated by the condition that the 
total luminosity carried by the accretion columns, $L_{col}$, is
a fraction of the accretion luminosity. This
fraction is determined by the disk truncation radius, $R_i$,
through $L_{col} =  \curf \times f 4 \pi R_*^2 = \zeta L_{acc}$,
with $L_{acc} = G \mdot M_*/R_*$, and
$\zeta = 1 - R_*/R_i$ (CG98).  
In the calculations shown below,
we take $R_i = 2.5 R_*$, consistent
with the magnetospheric infall models (section 2).

\begin{figure}
\plotone{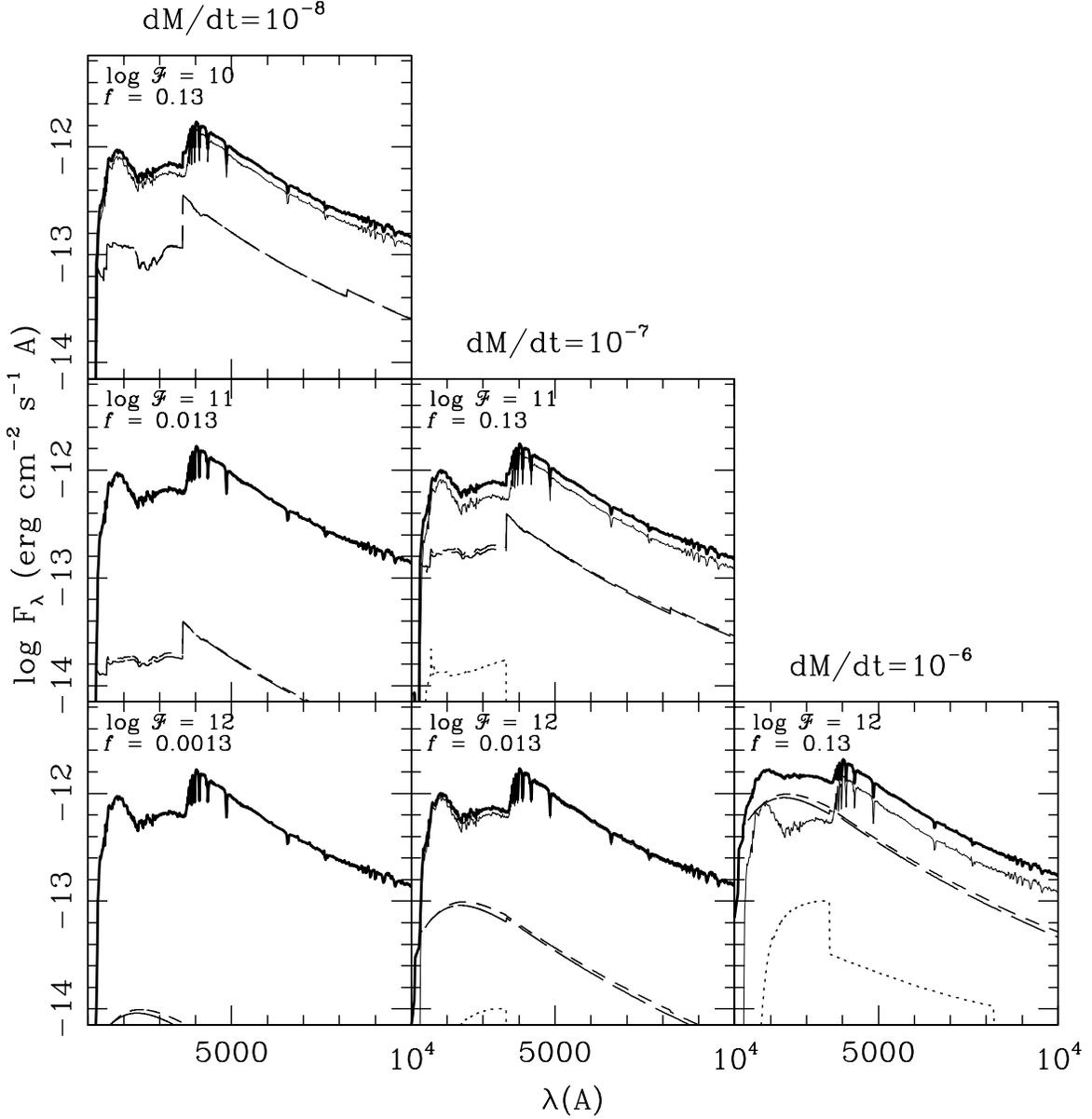}
\caption{Emergent fluxes for A2 photospheres with
mass accretion rates $10^{-8} \; \msunyr$ (left column),
$10^{-7} \; \msunyr$ (middle column), and
$10^{-6} \; \msunyr$ (right column). The energy flux of the
accretion column is $10^{10} \; \ecs$ (upper row),
$10^{11} \; \ecs$ (middle row), and
$10^{12} \; \ecs$ (bottom row). Filling factors for
the accretion columns $f$ are indicated in each panel.
Heavy solid lines indicate total fluxes, 
light solid lines correspond to fluxes
from the undisturbed photosphere, and dashed lines
to the emission of the accretion column. The contributions
from the shock regions are also indicated: heated photosphere
(dot-dashed) and preshock (dotted). See \S 2.2.
}
\label{shockflux}
\end{figure}

Figure \ref{shockflux}  shows the effect of the
accretion shock on the spectrum of
an underlying photosphere for an A2 star,
with mass and radius $M_* = 3 \; \msun$ and $R_* = 3 \; \rsun$.
Each row of panels corresponds to a different column energy
flux, with  $\curf$ increasing from $10^{10} \; \ecs$
at the top to $10^{12} \; \ecs$ at the bottom.
Each column corresponds to a different mass
accretion rate onto the star, from $10^{-8} \; \msunyr$
at the left, to $10^{-6} \; \msunyr$ at the right.
For a given mass accretion rate and energy flux,
the accretion columns must have
surface filling factors indicated in each panel
in Figure \ref{shockflux}.
Since $L_{acc} \propto \curf \times f$,
for a given accretion luminosity,
a lower energy flux results in a larger
surface coverage. Since $f$ cannot be $>1$,
high mass accretion rates cannot be achieved with low
$\curf$ columns alone.

The contribution to the 
flux from the different zones in the accretion
column, namely, the heated photosphere and
the preshock region, are indicated in Figure \ref{shockflux},
which also shows the emission from the undisturbed photosphere.
The photospheric fluxes are taken from the stellar population library of
Bruzual and Charlot (1993), while the fluxes from the heated
photospheres are calculated here, assuming
LTE and opacity sources as in CG98.
This calculation
does not include the hydrogen lines,
which in a real spectrum pile up at the longward
edge of the Balmer discontinuity, resulting
in an effective shift of the wavelength of the
jump, not present in the pure-continuum spectra
of the heated photosphere.

The main contribution to the flux in the visible and near-UV
in Figure \ref{shockflux} comes from the optically thick heated photosphere
below the shock.  The radiative flux emerging per unit area from this region
is $F^{hp} = F^s + F^*$, where
$F^s$ is the reflected irradiating flux from the shock
regions above the  optically
thick photosphere,
and $F^*$ the intrinsic stellar flux, 
entering the photosphere from below (CG98).
In turn, $F^s \sim 3/4 \, \,  \curf$
and $F^*  = \sigma T_*^4 \sim 3.6 \times 10^{11} \, (T_*/9000)^4$;
thus, if the energy fluxes carried by the accretion columns
in $\haebe$ stars are similar to those in CTTSs,
$\curf \sim 10^{11} - 10^{12} \; \ecs$, then
the accretion energy input in the heated photosphere
is comparable to or smaller than the intrinsic stellar flux.

The similarity between the two fluxes entering the
heated photosphere from above and below
means that the effective temperature of the heated photosphere,
$T_{hp} = (F_{hp}/\sigma)^{1/4}$, is only slightly higher
than that of the undisturbed photosphere,
as shown in Figure \ref{shockflux}. 
Since $f$ decreases as $\curf$ increases
for a
fixed mass accretion rate, the
contribution to the emergent flux from the 
heated
photosphere, $f F_{hp}$ (cf. eq.(\ref{totalf})), decreases as $\curf$ increases.
Note that even if the emission from the heated photosphere
and the undisturbed photosphere are separated by
some kind of ``deveiling'' procedure, the emission
from the heated photosphere does not give a direct measurement
of the accretion energy, except for very high values of $\curf$.
This situation contrasts sharply from that in the cooler CTTSs, 
for which $F^* \sim 2 \times 10^{10} \, (T_*/4000)^4$, and
accretion dominates the column emission.

However, there is a difference between
the emission from the heated photosphere region 
and that from the
undisturbed photosphere.
The temperature gradient in regions with $\tau_{Ross} \le 1$
in the heated photosphere
is shallower than that in the undisturbed photosphere,
because of the extra heating from the shock above
it. As a consequence, absorption features 
formed in the heated photosphere
will become weaker, and even disappear for
high values of $\curf$, in which the heated photosphere
is nearly isothermal. 
So, for high $\mdot$ objects, with large $f$,
the absorption lines will become increasingly
veiled.
The weakening of absorption features can be seen
in Figure \ref{shockflux}  comparing the
Balmer jump (in absorption)
in  the heated photosphere for increasing values
of $\curf$. The Balmer jump becomes weaker,
and disappears for $\curf \sim 10^{12} \ecs$.
As a result, the Balmer jump in the total
emergent fluxes is increasingly filled
in as $\curf$ increases and the mass accretion rate, i.e. $f$,
increases.

\begin{figure} 
\plotone{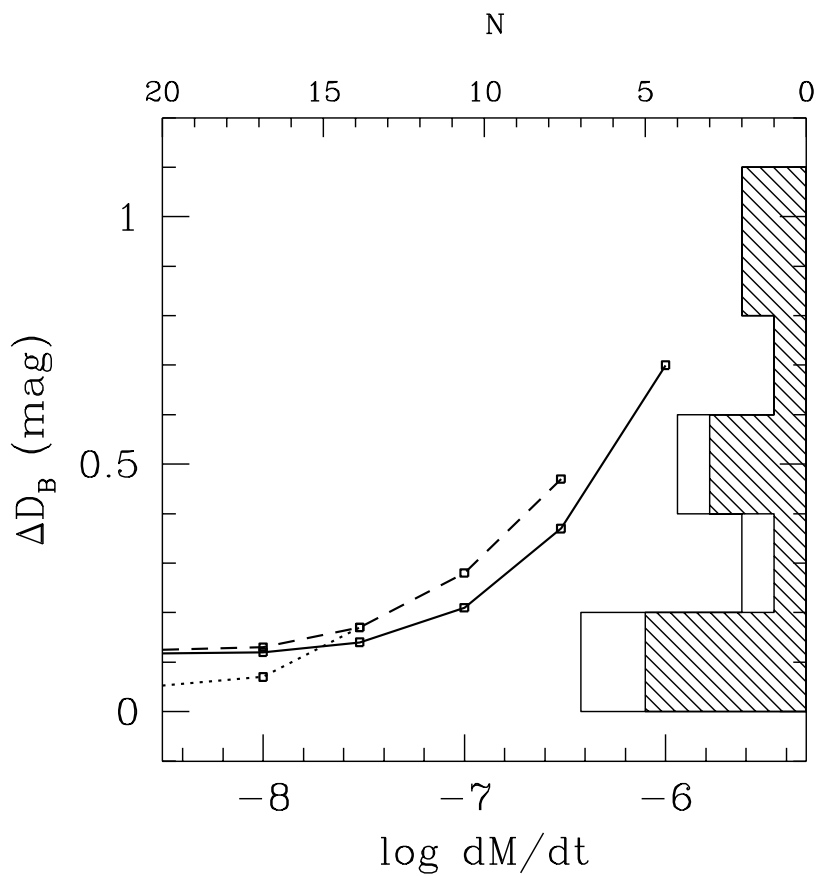}
\caption{Predicted excess in the Balmer discontinuity
as a function of mass accretion rate, for
different column energy fluxes:
$10^{10} \; \ecs$ (dotted line), 
$10^{11} \; \ecs$ (dashed line), and
$10^{12} \; \ecs$ (solid line). The distribution
of observed excesses for a sample of 16
HAe/Be stars from Garrison (1978) is shown at the right.
The shaded portion corresponds to the distribution
of excesses for the HAe stars in the Garrison sample.
Labels for the distribution are shown along the top axis.  
} 
\label{db}
\end{figure}

This property suggests that the strength of the
Balmer jump may provide a measure of the mass accretion
rate in HAe/Be stars. Unfortunately, this is the least
well-measured property of these stars. Garrison (1978, G78) 
carried out narrow (40 {\AA}) band photometry from 3400 to 8300 {\AA}
for a sample of HAe/Be stars, and noted that the Balmer discontinuity
was smaller than in main sequence stars of the same
spectral type. G78 defined a 
measurement of the strength of 
the Balmer discontinuity, ${\rm D_B}$,
as the difference in the magnitudes at both
sides of the discontinuity, and a measurement
of the {\it excess emission}, $\Delta {\rm D_B}$, as the difference
of $D_B$ for a given star to that of a standard
of the same spectral type.  
Figure~\ref{db}  shows 
the distribution of values of $\Delta {\rm D_B}$
for stars with $D_B$ in Garrison's sample.
We plot two histograms: one for the total sample of 16 stars
\footnote{We have excluded Z CMa, which is an FU Orionis object
(Hartmann \& Kenyon 1996).}; the other for the subsample of 12 stars
with spectral types B5 - A5.  The two distributions are essentially
the same.  Figure~\ref{db} also shows the predicted
excess in the Balmer discontinuity 
for shock models with mass accretion rates
between $10^{-9} \; \msunyr$ and $10^{-6} \; \msunyr$,
and energy fluxes $10^{10} \; \ecs$ to $10^{12} \; \ecs$.
We have calculated ${\rm D_B}$ from the ratio of fluxes
at 4000{\AA} and at 3640{\AA}, to make it consistent
with Garrison's measurements.
The Balmer jump excesses indicate low mass accretion rates
$\le 10^{-7} \; \msunyr$ for
most of the stars in the sample.
As stressed by G78,
$\Delta {\rm D_B}$ is a reddening
independent indicator, a property specially important
for the ``UX Ori'' stars, affected by obscuration
events in which it is difficult to disentangle
absorbed and scattered light (cf. Natta \& Whitney 2000).


A few of the stars in Figure \ref{db}  have Balmer
jump excesses consistent with high mass accretion
rates, $> 10^{-6} \; \msunyr$. For these stars, 
G78 finds excess fluxes relative to
the photosphere beyond $\lambda \sim 6500${\AA}.
Our models predict excesses at long wavelengths, 
due to the optically thin emission of the
pre-shock region, but the excesses are small.
For instance, a model with
$\mdot \sim 10^{-6} \; \msunyr$ has $\sim$ 0.1 mag
excess in $J-H$ and 0.15 mag excess in $H-K$.
Such small excesses would be difficult to disentangle
from observational errors and uncertainties due
to spectral type mismatches, as well as excess emission from the disk.

As discussed, spectra of HAe/Be stars with high mass accretion rates
should have weaker, or ``veiled'',
absorption lines than standard stars of the same spectral
type. The abnormally weak features would be formed in the
upper layers of the heated photosphere, and each line
could be affected differently, depending on its
exact height of formation. This is different from
the case of low mass stars, where a ``veiling continuum''
affects all photospheric lines in a given wavelength interval, and
can be easily extracted by subtraction of a standard
spectrum from the target.
For large $\mdot$ in HAe/Be stars, one would expect a large scattering 
in the values of the spectral types determined
using different absorption
features. Valenti, Johns-Krull, \& Linsky (2000), in a study 
of the IUE short wavelength spectra of a sample of 74 HAe/Be stars,
find that $\sim$ 70\% of the sample have spectral
features in the ultraviolet consistent with those
determined from optical spectra. 
In the case of UX Ori, discussed in section 2,
the optical spectrum of UX Ori shows that only
H$\alpha$ is in emission, and there is no veiling in the absorption lines;
analysis of the spectra as in Hernandez et al. (2002)
yields a spectral type
of A2 $\pm$ 1 subclass, using as indicators the Ca II K line, Fe II
lines, Ca I lines, and H$\gamma$.
On the other hand, Grady et al. (1995) find that the 1560{\AA} photoionization
edge in their IUE minimum light spectrum of UX Ori is consistent with
spectral types A1-A2.
The consistency of spectral type indicators
from the optical to the ultraviolet stresses
the lack of veiling in the absorption lines
and indicates that the mass accretion rate is
$< 10^{-7} \; \msunyr$.

In addition to continuum, the accretion column 
is expected to have and emission spectrum of
highly ionized metals 
(for instance, Si IV 1394, Si IV 1403, 
C IV 1549, C III 1909). These lines are observed in 
HAe/Be stars with spectral types 
later than A2, for which the weak photospheric
emission below 1600 {\AA} does not obliterate the lines
(Valenti et al. 2000); their strengths seem to be
comparable to those of CTTSs. For hotter stars, these
lines may appear in absorption, and even have a
wind component observed at high resolution (Praderie et al. 1986; Bohm et al. 1996;
Bouret et al. 1997; Grady et al. 1999). The wind
may also be responsible for spectral
features that appear deeper than expected from their spectral types
in some regions of low resolution ultraviolet spectra
(mostly corresponding to blends of Fe II lines,
Valenti et al. 2000).
In particular, we note that free-free emission from the wind may be
responsible for the continuum excess between $\sim$ 6500 {\AA} and
$\sim 1 \mu$m in the stars with large Balmer jump excesses (Garrison
1978). Neither present shock models (see Fig.~\ref{shockflux})
nor wall emission
(see below, section 3) can explain this excess. Appropiate modeling of
the wind region is required to test this hypothesis further, and
yield measurements of mass loss rates, thus providing lower limits to
the mass accretion rates.

\section{Accretion disk models}

Having determined that disks are accreting in $\haebe$ stars,
we now examine the structure of the inner disks in order to address
another question posed in the introduction, namely, whether or not
they are optically thin, so that (1) they
do not produce a large flux excess over the
photosphere at near infrared wavelengths, and
(2) they allow stellar radiation to reach the
inner wall postulated by Natta et al (2001), Tuthill, Monnier \&
Danchi (2001) and DDN01 to explain
the near-IR photometric excess and interferometric measurements.
To illustrate the issues involved, we have calculated pure gaseous
models for the inner disk regions of a typical $\haebe$ system,
with assumptions as those of D'Alessio \etal (1999, 2001).

Optically-thin dust cannot be present interior to the dust destruction
radius, $R_d$, due to the radiation field of the central object.  In principle
dust could exist if it were ``shielded'' from the stellar radiation
by gas opacity, but in such a case the ``irradiation surface'', defined
as the surface at which the optical depth to the star (at characteristic
frequencies) is $\sim 1$, has to be determined by the gas opacity.
Thus any dust present
at $R < R_d$ would have to be below this surface.  We have simplified
the already complex treatment of this gaseous layer (see below) 
by assuming that dust is not present anywhere interior to the irradiation
surface.

If the sublimation temperature for silicates is taken as $\sim$ 1500 K
and the dust wall receives radiation frontally,
as  proposed by Natta et al. (2001) and DDN01, then the wall
would be at a radius $\sim 0.5$ AU for the effective temperature 
and radius of a typical Ae star (see DDN01).
The inner gaseous disk then corresponds
to the region between the magnetospheric radius
$R_{mag} \lesssim 3 \ R_* \sim 0.04$ AU and $\sim$ 0.5 AU.
We calculate pure gaseous $\alpha$ irradiated accretion disk
models for this region, using the typical $\haebe$ stellar parameters.

The vertical temperature structure is calculated
using the formulation described by Calvet et al. (1991, 1992),
modified to include viscous dissipation, which is quantified
using the $\alpha$ prescription (Shakura \& Sunyaev 1972).
The  stellar radiation field is treated as
a parallel beam impinging on the disk surface at an angle
cos$^{-1} \ \mu_0$ relative to the normal. A fraction of the energy in
the incident beam is scattered,
maintaining the same frequency range as the stellar radiation field.
The other fraction of the incident energy is absorbed by disk material,
and we assume it is re-emitted at a frequency range characteristic of the
local temperature.
In the Calvet et al. (1991, 1992) formulation, the interaction
between the direct and scattered stellar and disk radiation
fields and the material is described by mean opacities,
which we calculate self-consistently from monochromatic
opacities using relevant sources for the
conditions of the inner gaseous disk.

For an $\haebe$ star, approximately half of the stellar radiation is
emitted at wavelengths shorter than $\sim$ 0.45 $\mu$m, in the UV.
In this range,
we considered absorption due to H$_2$ (B-X, C$^+$-X and C$^-$-X,
Abgrall et al. 1993) and CO (fourth positive system,
Kurucz 1976) transitions, calculated with LTE molecular abundances.
We found that the opacity due to the superposition
of the rotational lines of CO and H$_2$ is so large for $\lambda \lesssim 0.2 \ \mu m$ 
that most of the stellar UV radiation
is  absorbed in the uppermost layers of the disk.
A fraction of this UV radiation (with $\lambda < $ 0.11 $\mu$m) 
will photodissociate H$_2$ and CO (see for instance van Zadelhoff et al. 2003).
Another fraction will 
be absorbed in CO and H$_2$ lines below the photodissociated layer.
To calculate the structure of the gaseous inner disk,
we make the crude approximation that the incident stellar radiation
with $\lambda < 0.2 \ \mu m$ does not penetrate and heat the disk below a
geometrically thin upper layer.  In other words,
we assume  that all this radiation is absorbed and/or scattered
by photodissociated molecules, or scattered in high-opacity lines
that become saturated at the uppermost layers of the disk atmosphere.

The remaining stellar radiation at $\lambda > 0.2 \ \mu m$ is
mostly absorbed by TiO.  We have incorporated the opacities of the 
$\alpha$, $\beta$, $\gamma$, $\gamma^\prime$, $\delta$,  $\epsilon$,
and $\phi$ systems (J\"orgensen 1994).
We also included the opacity due to CN and CH (Kurucz 1995), although it
turned out to be unimportant in the disk temperature range
($\sim 500-3000$ K).
At the characteristic disk wavelengths, the most important opacity sources are
red and infrared bands of H$_2$O (Auman 1967), pure-rotational bands 
of H$_2$O and OH (Tsuji 1966), and transitions of the
electronic ground state of CO (Kirby-Docken \& Liu 1978).
We assume LTE and calculate the populations of TiO, H$_2$O, CO$_2$,
C$_2$, N$_2$, O$_2$, CH, CN, CO, NH, NO, OH, H$_2$, H$_2^+$, H$^-$, H$^0$
using the method described by Mihalas (1967).
The scattering of stellar radiation is mostly due to Rayleigh
scattering of H$_2$, with smaller contributions from
Rayleigh scattering of H, He (Dalgarno 1962)
and electronic scattering.

Even with all these opacity sources, there is a lack of absorption 
in the wavelength range between $\sim 0.2-0.45 \ \mu m$. In this 
interval, radiation interacts with the gas only through 
Rayleigh scattering. Since scattering processes do not contribute 
to the disk heating, we make the additional approximation that only 
half of the incident stellar
flux (mostly with $\lambda > 0.45 \ \mu m$ or in the gaps between lines) 
contributes to the 
heating of the  
interior of the disk.

The temperature as a function of the Rosseland mean optical depth $\tau_R$,
the mean scattering coefficient, and the ratio between the opacity
to the stellar radiation and the Rosseland mean opacity, $q$,
are calculated at each height and radius
using equation (14) from Calvet et al. (1991),
with a simultaneous solution of the 
hydrostatic equilibrium equation in the vertical direction.
In order to evaluate the mean scattering coefficient and $q$ 
(both assumed to be constant through 
the disk atmosphere), we computed
the opacities at the upper temperature $T_0$ and at a constant 
pressure $P_0=10^{-6}$  dyn/cm$^2$, assuming both to be representative 
values for the disk upper atmosphere.
Given the proximity of the gaseous inner disk zone
to the central star ($R \lesssim 40 \ R_*$), the curvature of the
irradiation surface is not an important factor in the calculation of
the stellar irradiation.
So, we take the irradiation flux $F_{irr}$  and incident angle 
as those of a flat disk, 
\be
F_{irr} \approx  {1 \over 2} \sigma_R T_*^4 \biggl ( {R_* \over R} \biggr)^2 \mu_0,
\en
with
\be
\mu_0 = {2 \over 3 \pi} \biggl ( {R_* \over R} \biggr),
\en
the cosine of the angle between the incoming stellar beam and the normal
to the disk. The factor $1/2$ is introduced to account for 
our assumption that only half of the stellar radiation heats the regions
of the disk below the uppermost layer (i.e., only half of the stellar radiation
has $\lambda > 0.45 \mu$m). 

\begin{figure}
\plotone{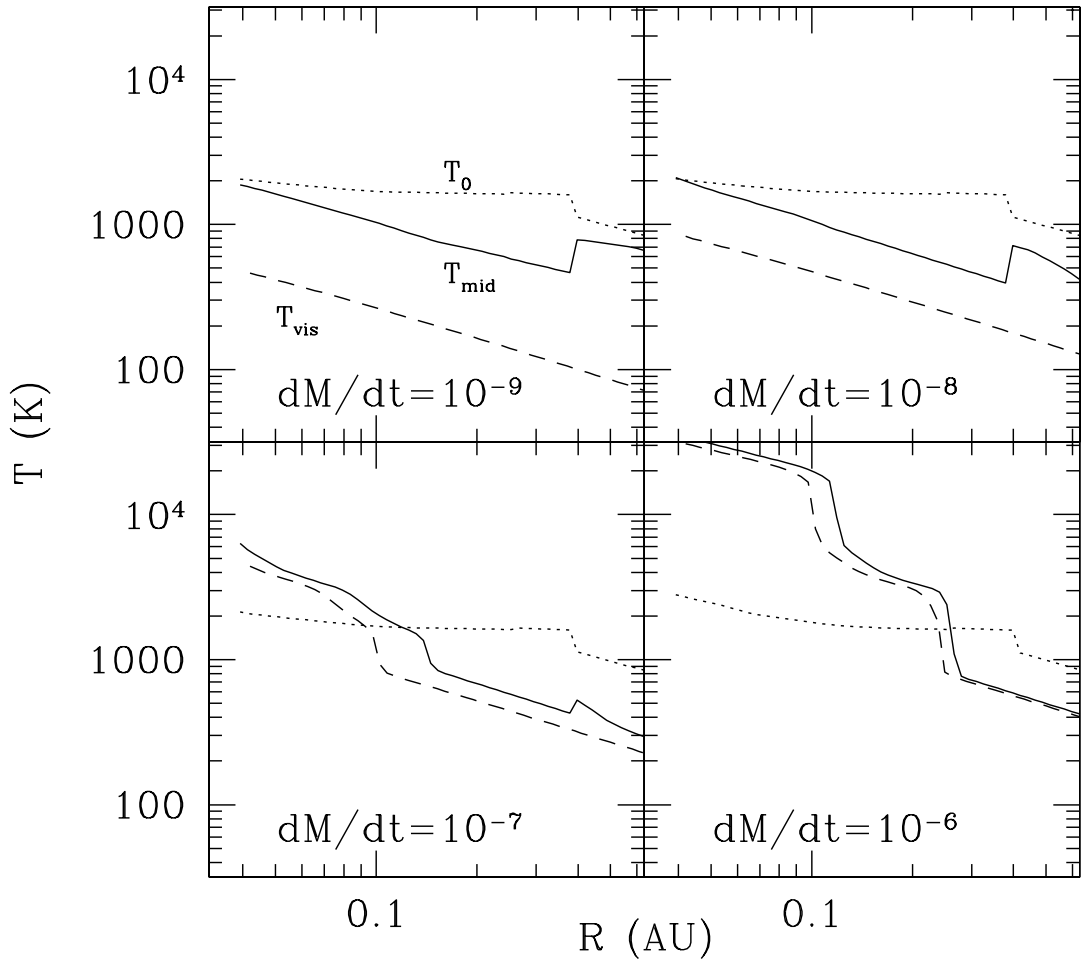}
\caption{Characteristic disk temperatures for mass
accretion rates $10^{-9}, 10^{-8}, 10^{-7}$, and $10^{-6} \; \msunyr$,
indicated in each panel: upper layer temperature $T_0$
(dotted line), midplane temperature $T_{mid}$ (solid line),
and viscous temperature $T_{vis}$ (dashed line).
}
\label{estructura}
\end{figure}

With these assumptions and procedures,
we calculate gaseous disk models for  mass accretion rates,
$\mdot=10^{-9}, \ 10^{-8}, \ 10^{-7}$ and $10^{-6} \ \msunyr$, assuming a
constant viscosity parameter $\alpha=0.01$. 
The resultant temperature structures of these models
are shown in Figure~\ref{estructura}.
As shown by the dotted lines in
Figure~\ref{estructura}, the optically thin upper atmosphere
has a temperature around  $\sim$ 2000~K, 
for all values of $\mdot$.
This is the equilibrium temperature corresponding to
radiative heating by stellar radiation with $\lambda > 0.45 \ \mu m$ and
radiative cooling, for a constant pressure of $10^{-6}$ dyn/cm$^2$.  
This nearly constant temperature is due to the fact that 
in this temperature range for the assumed constant
pressure, the cooling increases with temperature much faster than the heating,
producing a thermostatic effect.  
Although the upper layer temperature is similar to or lower than the condensation
limit for silicates, there cannot be dust in these
(optically thin) regions because a dust grain would be heated by the stellar
radiation to a temperature above its sublimation limit.
As we have already mentioned, dust could exist at deeper layers if it were shielded
by upper, optically thick gaseous layers;
however, here we assume that the inner disk is free
of dust at every height. 

The Rosseland mean optical depth of the gaseous inner disk
increases with $\mdot$. For $\mdot=10^{-9} \ \msunyr$,
the disk is optically thin, and the midplane temperature
is similar to or lower than the upper
atmospheric temperature, as shown in Figure \ref{estructura}.
At higher mass accretion rates, the innermost parts of the 
disk become increasingly optically thick. The radius inside which the disk
is optically thick to its own radiation increases
with $\mdot$, reaching 0.25 AU for $\mdot=10^{-6} \ \msunyr$.
As a result, the midplane  temperature increases with 
$\mdot$ as shown in Figure \ref{estructura}.
For $\mdot \gtrsim 10^{-6} \ \msunyr$, it becomes higher
than $T \sim 10^4$ K in the innermost regions,
corresponding to the thermally unstable
regime where hydrogen is partially ionized and
the dominant opacity source is $H^{-}$ (Bell \& Lin 1994).

\begin{figure} 
\plotone{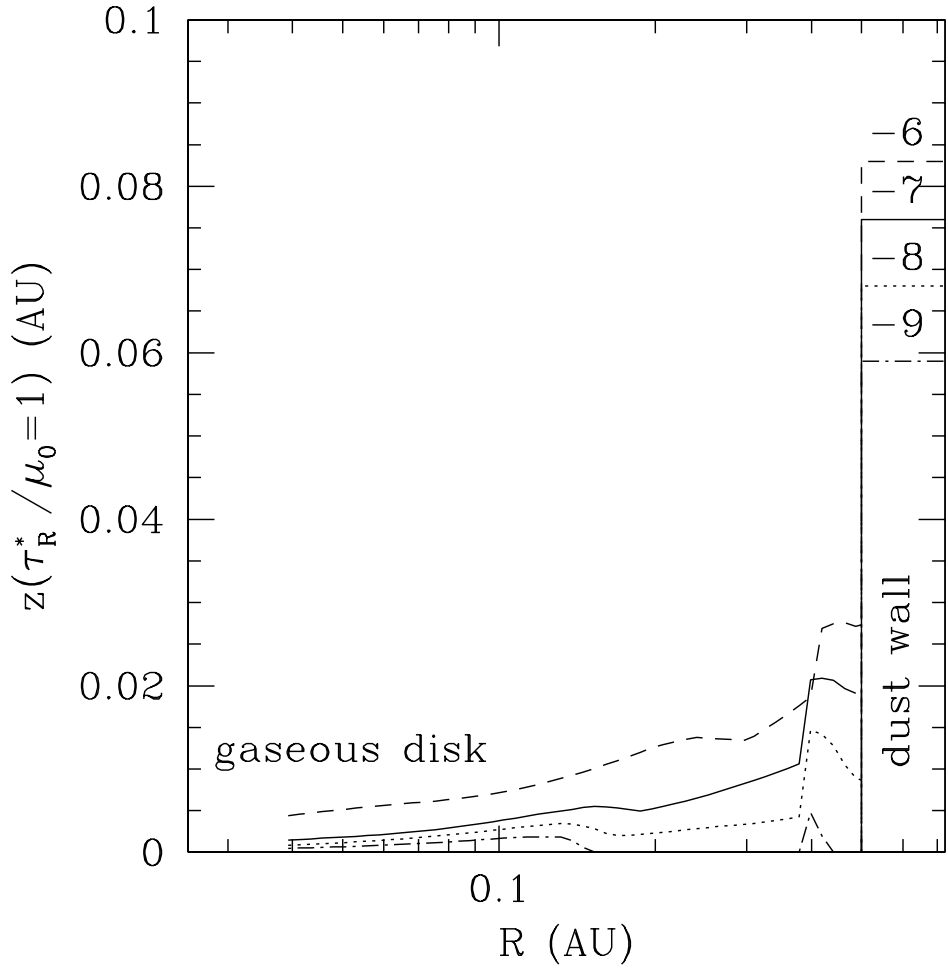}
\caption{Height of the inner gaseous disk surface
for mass 
accretion rates 
$10^{-9} \; \msunyr$ (dash-dot), 
$10^{-8} \; \msunyr$ (dot), 
$10^{-7} \; \msunyr$ (solid), 
$10^{-6} \; \msunyr$ (dash). 
The height of the dust wall is indicated
by a solid square (see \S 3).
} 
\label{zetas} 
\end{figure}

Next, we calculate the height of the surface
above the midplane, where the mean optical depth to 
stellar radiation is unity, $z_s$.
In our previous modeling efforts with dusty disks, we have
used the Planck mean optical depth to estimate $z_s$;
in the present case, the Planck mean optical depth to the stellar radiation
(using the incidence angle for a perfectly flat disk)
is larger than $10^3$, even for a low mass accretion rate
like  $\mdot=10^{-9} \msunyr$. 
However, the Planck mean opacity is not representative of the
mean opacity for these cool gaseous disks; since the 
absorption of radiation is mostly
due to a myriad of molecular rotational lines,
stellar radiation is transported through the disk 
in the opacity gaps between lines.
Thus, we have used a harmonic mean similar to the
Rosseland mean, which gives a better
representation of the actual transport of
radiation. Figure \ref{zetas} shows the height of the
surface for models in Figure \ref{estructura},
calculated with the harmonic mean opacity
to the stellar radiation.
We find that 
for $\mdot=10^{-9}  \ \msunyr$, the
harmonic mean optical depth to the
stellar radiation is less than 1 for $R > 0.1$ AU.
At $R \sim 0.5$ AU,
$z_s^{gas} \sim$ 0.008, 0.02, 0.03  AU
for $\mdot\,  =\,  10^{-8},\, 10^{-7},\, 10^{-6} \ \msunyr$, respectively. 
This corresponds to $z_s / H(T_c) \sim$ 1, 2.6, 3.3, 
where $H(T_c)$ is the scale height evaluated at the midplane
temperature of each of these disk models
(cf. Figure \ref{estructura}).

At the dust condensation radius, the dust becomes the most important
opacity source. For comparison with the gaseous disk, we estimate the 
height where the radial optical depth to the stellar radiation 
is one, behind the dusty wall. We assume that the gaseous inner disk 
does not absorb stellar radiation in the path
between the star and the wall significantly (confirmed below).
To calculate the optical depth to the stellar radiation in a disk annulus 
of radial thickness $\Delta R$, we assume that the annulus is isothermal 
with a temperature $T_w \approx $1500 K. Thus, the  density can be 
approximated by a Gaussian $\rho = \rho_c \exp(-z^2/2H^2)$, 
where the midplane density is $\rho_c \approx  \Sigma(R)/ \sqrt{2 \pi} H(R)$, 
the surface density is $\Sigma(R) \approx \Md/3 \pi \alpha c_s H(R)$,
the scale height is  $H(R)=c_s/\Omega$, 
the sound speed is $c_s = \sqrt{k T_w/\mu m_H}$, and $\Omega$ is the 
Keplerian angular velocity.
The optical depth to the stellar radiation is 

\be
\tau_{rad}^* = \int_r^{r+\Delta r} \rho(r) \, \chi_* \, dr, 
\en
where $r$ is the radial distance from the star to a point $(R,z)$ in the disk, 
such that  $dr = dR \sqrt{R^2+z^2}/R$, and $\chi_*$ is the total
mean dust opacity to the stellar radiation.
Assuming that $\Sigma$, $H$, and $\Omega$ are 
constant from $R$ to $R+\Delta R$, the integral can
be approximated by 

\be
\tau_{rad}^* \approx \chi_*  <\rho>\left ( R ^2 + z^2  \right )^{1/2} {\Delta R \over R}.
\en
The mean density  $<\rho>$ can be estimated as an average of the densities 
at $(R,z)$ and at $(R+\Delta R, z+\Delta z)$, where $\Delta z \approx 
\Delta R \ z/R$, i.e.,  
\be
<\rho> \, \approx {\Sigma(R) \over \sqrt{8 \pi}H(R) }\biggl [ e^{-z^2/2H^2}+ e^{-z^2(1+\Delta R/R)^2/2H^2} \biggr ]
\en
With $\Delta R/R \approx 0.1$, and estimating $\chi_* $ by assuming
that the dusty wall contains grains with 
a maximum size $a_{max}=0.1 \  \mu m$,
the heights where $\tau_{rad}^*\approx 1$ are 
$z_s^{dust}/ H(T_{w}) \sim $ 3.8, 4.3, 4.8 and 5.3, corresponding to 
$\mdot\,  =\,10^{-9},\,  10^{-8},\, 10^{-7},\, 10^{-6} \ \msunyr$, 
respectively. These values are roughly consistent with the value 
found by DDN01, 
$z_s^{dust} \sim 5 H(T_{w})$, assuming  
an arbitrary power
law surface density 
for the disk.

The temperature of the dust wall 
$T_{w}$  is about 2 to 3 times higher than
the temperature of the gas disk at the same radius
(cf. Figure~\ref{estructura}).
Thus, $z_s^{dust} / z_s^{gas} \sim$  10, 4, 3 for 
$\mdot\,  =\,  10^{-8},\, 10^{-7},\, 10^{-6} \ \msunyr$,
as indicated in Figure \ref{zetas}.
The fraction of the surface of the dust wall that
receives direct stellar radiation is $\sim$ 1, 0.9, 0.8, 0.7
for $\mdot\,  =\,  10^{-9},\,10^{-8},\, 10^{-7},\, 10^{-6} \ \msunyr$,
and the portion of the wall in contact with the
optically thick (to the stellar radiation) gaseous disk
is small.  Thus, at least for $\mdot \lesssim 10^{-7} \ \msunyr$, 
allowance for the stellar flux that cannot reach the wall because of
absorption in the gaseous inner disk
represents only a small correction to the Natta et al. (2001) and DDN01 treatments.

\begin{figure}
\plotone{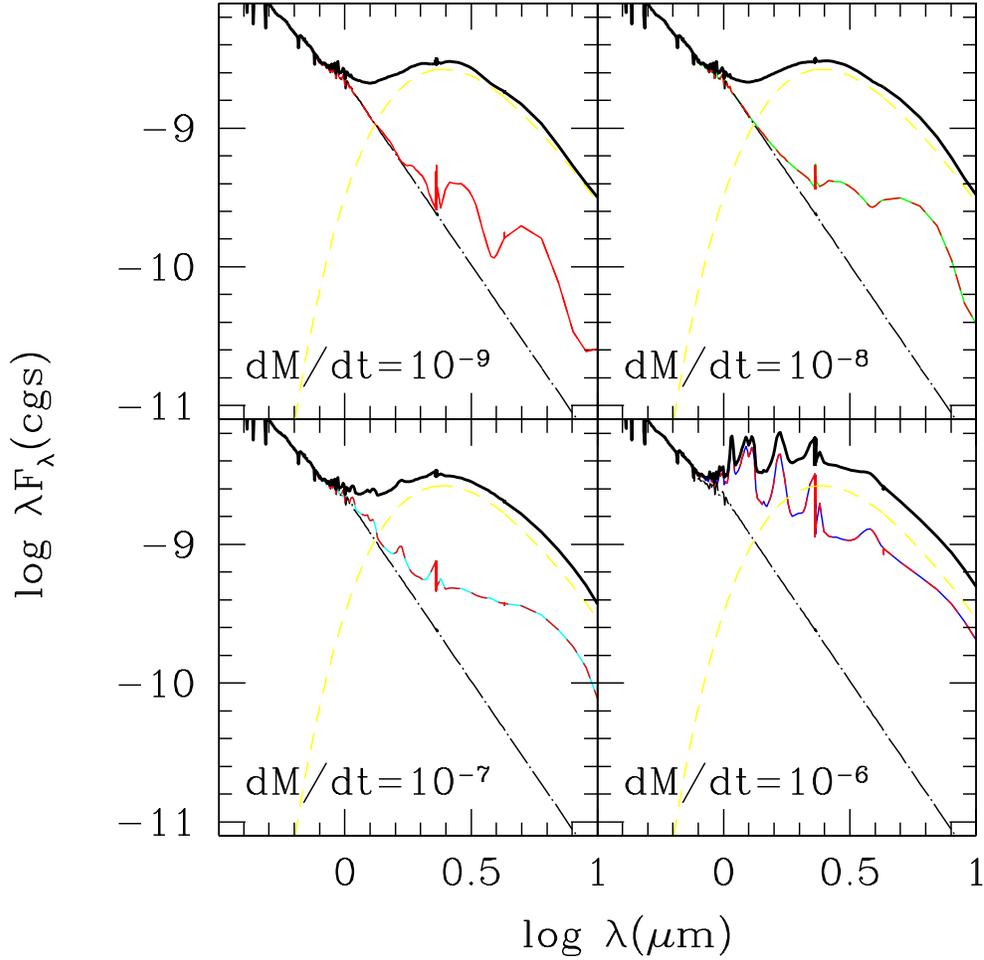}
\caption{Emission from the inner gaseous disk
(solid line) for mass
accretion rates $10^{-9}, 10^{-6}, 10^{-7}$, and $10^{-6} \; \msunyr$,
indicated in each panel. The disks are seen pole-on
and the fluxes are calculated at 440 pc.
Emission by the dust wall, indicated by the dashed line,
is shown as a blackbody at $T$ = 1500K,
covering a solid angle $\sim$ 250 times the stellar solid angle,
which is representative of observed spectra (see HSVK92). 
The SED of the stellar photosphere (dash-dotted line) was taken from the
Bruzual \& Charlot (1993) spectral library.  The total flux is shown by the
heavy solid line. The outer disk does not contribute at
near-IR wavelengths (cf. Natta et al 2001).
}
\label{diskflux}
\end{figure}

The flux from the inner gaseous disk, in the most optimistic case
of a pole-on orientation, is shown
in Figure~\ref{diskflux}.  We also include the contribution
from the wall, which we have estimated with a
blackbody of temperature 1500 K and an emitting
area covering a solid angle 250 times the stellar
solid angle (as expected for a vertical wall with height
$z_s \sim  0.08$ AU, seen almost edge-on but without occultation
of the emitting region).  This blackbody contribution roughly matches
typical HAe spectra (see HSVK92).
The disk emission depends on $\mdot$;
in particular, the total emergent flux from 
the gaseous inner region in the $K$-band, relative to the star,
is $F_{K} / F_{*,K} \sim 0.5, 0.6, 1, 4 $
for $\mdot\,  =\,  10^{-8},\, 10^{-7},\, 10^{-6} \ \msunyr$.
The inner gaseous zones of disks with $\mdot \gtrsim 10^{-6} \ \msunyr$
are optically thick, and emit a large continuum excess 
in the near-IR, in agreement with Hartmann et al. (1993).
For $\mdot \lesssim 10^{-7} \ \msunyr$, the contribution
to the total flux from the inner gaseous disk is negligible.
The inner disk emits rotational-vibrational CO bands around 2.3 $\mu m$
for $\mdot \lesssim 10^{-7} \  \msunyr$;
however, these CO emission bands are obliterated
by the continuum emerging from the outer disk and the wall,
in agreement with observed near-IR spectra of Herbig Ae/Be stars
(unpublished IRTF SpeX data; Ishii et al. 2003).


\section{Discussion}

\begin{figure}
\plottwo{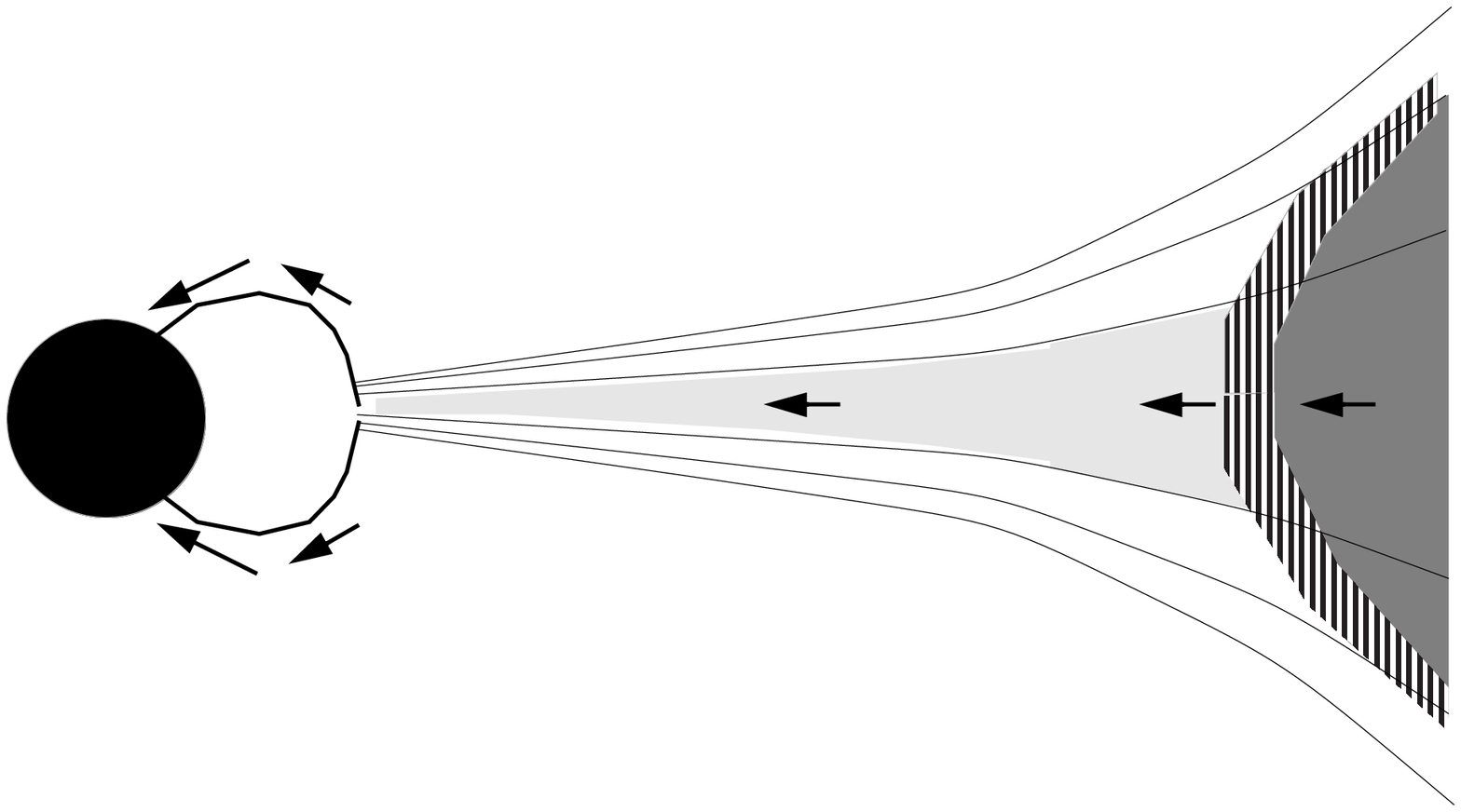}{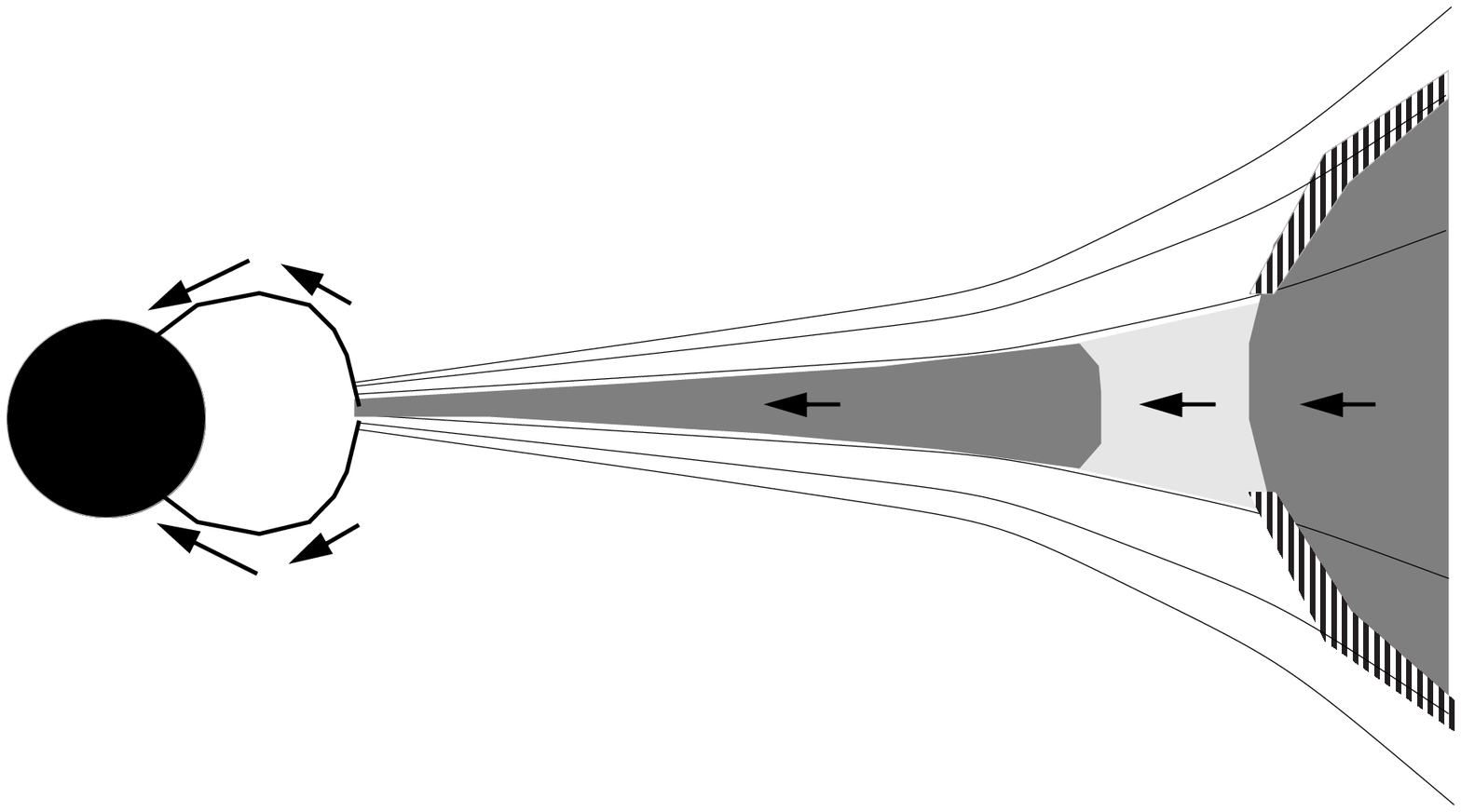}
\caption{A schematic view of disk accretion in HAeBe stars.
Lines of constant gas density in the disk are indicated by thin
curves; regions of relatively low (gaseous) optical depth are indicated
by light shading, while optically-thick regions are shown by dark shading.
At low accretion rates, the inner disk is optically thin (left panel),
allowing direct irradiation of the dust ``wall'' (hatched shading).
At higher accretion
rates, even if the inner disk is optically thick (right panel), it is
geometrically thinner than the dust ``wall'', again allowing for direct
(normal incidence) irradiation.  In this way the DDN01 model can be
maintained while allowing for an inner disk which feeds magnetospheric
accretion columns, funneling material down onto the central star (arrows).
}
\label{disk_cartoon}
\end{figure}

We have shown that the magnetospheric accretion model can account for the overall
line profiles of UX Ori and, by extension, similar systems.  This
requires substantial amounts of accreting gas which lie well within the
radius at which dust is destroyed.  Although we presume that the accreting
gas departs from a disk structure within a few stellar radii of the photosphere,
accounting for the large radial velocities observed, the co-rotation radius
within which such magnetospheric accretion can occur is well interior to
the dust destruction radius.  Therefore, there must be a gaseous accretion
disk between the dust opacity-dominated outer disk, responsible for most of
the infrared excess emission, and the star (see Fig.~\ref{disk_cartoon}).
We have shown that for plausible
parameters, this inner accretion disk can be optically thin.
Even at somewhat higher accretion rates, an optically-thick inner disk can
be geometrically thin, and thus not block most of the
incident stellar radiation from reaching the outer dust disk
This is consistent with the basic picture of Natta et al. (2000) and DDN01,
in which there is a disk ``wall" where dust sublimates.

Our model, in which an accreting gas disk extends interior to the
dust destruction radius, may help address or reduce stability problems.
As pointed out by DDN01, the rapid increase in scale height of their model 
implies an inward radial gas pressure force at the dust destruction radius.
DDN01 suggest that this might be countered by having the inner regions rotating 
faster than Keplerian, but this seems unlikely.  The problem is alleviated to some
extent in our model, in which there is an inner gas disk; the scale height variation
is less than the photospheric surface $z_s$ variation.   

Our claim of magnetospheric accretion in UX Ori-type Herbig Ae/Be stars
leads to the question of where the requisite magnetic activity originates.
This is still far from clear, as the motivation for theoretical investigation
has been lacking.  Strong magnetic fields on the order of several kG have been
measured in the lower-mass T Tauri stars (e.g. Johns-Krull, Valenti, \&
Koresko 1999).  However, similar measurements for HAe/Be stars may prove
exceedingly difficult, due to the significant rotational broadening of
photospheric absorption lines.  Given the parameters of our best
line profile model, and using the relations given in K\"onigl (1991),
we estimate a possible magnetic field strength of $\sim$200 G (roughly
half that if the small magnetospheric size is used) for UX Ori.
We point out that at a younger age, further
back along their radiative tracks, HAe stars once had cooler photospheres,
and likely sustained convective dynamos (for example, intermediate-mass
T Tauri stars such as T Tau and SU Aur, which have clear accretion signatures,
may eventually become what we would classify as Herbig Ae stars).
Thus, their current magnetic activity may be a relic of this earlier time.
Detailed studies of the timescales of magnetic decay in such objects are
needed in order to investigate this hypothesis further.

%
%
This work was supported in part by NASA grant NAG5-9670.
The authors acknowledge useful discussions and assistance from 
Vladimir Escalante, Javier Ballesteros, Antonella Natta, and
Kees Dullemond.  PD acknowledges grants from DGAPA and CONACyT.

\clearpage 

\end{document}